\documentclass[aps,pra,twocolumn,longbibliography,superscriptaddress]{revtex4-2}
\usepackage{amsmath,amssymb,amsfonts}
\usepackage{color}
\usepackage{enumerate}
\usepackage{graphicx}
\usepackage{epstopdf}
\usepackage[]{hyperref}
\hypersetup{
    pdftitle={qfluid spectra},
    pdfauthor={A. S. Bradley},
    pdfsubject={gpe spectra},
    colorlinks=true,
    linkcolor=blue,
    citecolor=blue,
    filecolor=black,
    urlcolor=blue
}
\def\urlprefix{}
\def\url#1{}

\newcommand{\rr}{\mathbf{r}}

\newcommand{\del}{\partial}
\newcommand{\intr}{\int d^3\rr\;}
\newcommand{\ali}[1]{\begin{align}#1\end{align}}

\newcommand{\x}{{\mathbf x}}

\newcommand{\ip}[2]{\langle#1|#2\rangle}

\newcommand{\mbf}[1]{\mathbf{#1}}

\newcommand{\uu}{{\mathbf u}}

\newcommand{\vv}{{\mathbf v}}
\newcommand{\ww}{{\mathbf w}}

\newcommand{\kk}{\mbf{k}}

\newcommand{\FT}[1]{{\cal F}(#1)}
\newcommand{\IFT}[1]{{\cal F}^{-1}(#1)}

\newcommand{\eref}[1]{(\ref{#1})}
\newcommand{\eeref}[1]{Eq.~(\ref{#1})}
\newcommand{\sref}[1]{Section~\ref{#1}}
\newcommand{\aref}[1]{Appendix~\ref{#1}}
\newcommand{\fref}[1]{Fig.~\ref{#1}}

\DeclareMathOperator{\sinc}{sinc}

\begin{document}

\title{Spectral analysis for compressible quantum fluids}

\author{Ashton S. Bradley}
\affiliation{Department of Physics, University of Otago, Dunedin 9016, New Zealand}
\affiliation{Dodd-Walls Centre for Photonic and Quantum Technologies}
\author{R. Kishor Kumar}
\affiliation{Department of Physics, University of Otago, Dunedin 9016, New Zealand} 
\affiliation{Dodd-Walls Centre for Photonic and Quantum Technologies}
\author{Sukla Pal}
\affiliation{Department of Physics, University of Otago, Dunedin 9016, New Zealand}
\affiliation{Dodd-Walls Centre for Photonic and Quantum Technologies}
\author{Xiaoquan Yu}
\affiliation{Graduate School of  China Academy of Engineering Physics, Beijing 100193, China}
\affiliation{Department of Physics, University of Otago, Dunedin 9016, New Zealand}

\date{\today}

\begin{abstract}
Turbulent fluid dynamics typically involves excitations on many different length scales. Classical incompressible fluids can be cleanly represented in Fourier space enabling spectral analysis of energy cascades and other turbulence phenomena. In quantum fluids, additional phase information and singular behavior near vortex cores thwarts the direct extension of standard spectral techniques. We develop a formal and numerical spectral analysis for $U(1)$ symmetry-breaking quantum fluids suitable for analyzing turbulent flows, with specific application to the Gross-Pitaevskii fluid. Our analysis builds naturally on the canonical approach to spectral analysis of velocity fields in compressible quantum fluids, and establishes a clear correspondence between energy spectral densities, power spectral densities, and autocorrelation functions, applicable to energy residing in velocity, quantum pressure, interaction, and potential energy of the fluid. Our formulation includes all quantum phase information and also enables arbitrary resolution spectral analysis, a valuable feature for numerical analysis. A central vortex in a trapped planar Bose-Einstein condensate  provides an analytically tractable example with spectral features of interest in both the infrared and ultraviolet regimes. Sampled distributions modeling the dipole gas, plasma, and clustered regimes exhibit velocity correlation length increasing with vortex energy, consistent with known qualitative behavior across the vortex clustering transition. The spectral analysis of compressible quantum fluids presented here offers a rigorous tool for analysing quantum features of superfluid turbulence in atomic or polariton condensates.
\end{abstract}

\maketitle

\section{Introduction}
Many features of the complex dynamics of classical fluid turbulence are more clearly represented using a Fourier representation of the energy distribution across length scales. 
Spectral analysis provides an essential tool for understanding classical~\cite{kolmogorov_local_1941,grant_turbulence_1962} and quantum~\cite{nore_kolmogorov_1997} turbulence, where energy transport over a large range of scales is a central feature of steady-state turbulent flow, and provides insights into energy cascades~\cite{navon_emergence_2016,navon_synthetic_2019}, and non-thermal fixed points~\cite{karl_strongly_2017} in far from equilibrium systems. A clear formulation of spectral analysis is an important open problem in compressible quantum fluids~\cite{nore_kolmogorov_1997,bradley_energy_2012} where turbulent flows typically involve significant interaction between fluid velocity and density~\cite{griffin_energy_2022}. As perhaps the simplest example of such a fluid, dilute-gas Bose-Einstein condensates (BECs) exhibit rich phenomenology due to the presence of coupled incompressible (vortex) and compressible (phonon) excitations~\cite{berloff_scenario_2002,tsubota_quantum_2009}. Fluids with similar properties occur in a range of quantum degenerate systems including in unitary Fermi gases~\cite{hossain_rotating_2022-1}, Bose-Fermi
mixtures~\cite{yao_observation_2016}, and polariton condensates~\cite{koniakhin_2d_2020,panico_onset_2022}. Quantum fluids are distinguished by the central role of quantum phase information in their dynamics. However, a complete spectral formulation for compressible quantum fluids that includes all quantum phase information has remained an open problem. A number of important questions could be addressed by such a formulation: what is truly quantum in quantum turbulence? How does quantum phase information enter quantum turbulence? What is the physical role of enstrophy? 

Here we generalize the Helmholtz decomposition approach for velocity power specta~\cite{nore_kolmogorov_1997} to formulate a general spectral analysis for compressible quantum fluids described by a $U(1)$ symmetry breaking order parameter. Our formulation provides a natural definition of energy spectral densities that includes all quantum phase information, and reduce  to familiar velocity power spectra when this information is neglected. We formulate the general problem of computing a spectral density in terms of a spectral inner product that allows angular integrals in $k$-space to be carried out analytically, with obvious formal advantages. Numerically, a remaining spatial two-point correlation may be accurately evaluated using a discrete Fourier transform. Spectra can thus be calculated without binning approximation. The $k$-space accuracy also allows high resolution position correlations to be easily computed from spectra. Our approach is particularly useful for analyzing long-wavelength phenomena in planar quantum fluids where spectral condensation dominates at high vortex energy~\cite{billam_onsager-kraichnan_2014,simula_emergence_2014}, and offers a pathway to calculating energy fluxes in quantum turbulence. 

The Gross-Pitaevskii equation (GPE) provides an accurate description of trapped Bose-Einstein condensates (BECs) far below the BEC transition temperature~\cite{dalfovo_theory_1999}. As an exemplar of a $U(1)$ symmetry breaking quantum fluid dynamics, the GPE offers a testbed for spectral analysis of turbulent flows~\cite{nore_kolmogorov_1997,kobayashi_kolmogorov_2005,bradley_energy_2012,reeves_inverse_2013,curtis_characterizing_2019}, created in experiments by forced injection of quantum vortices or acoustic excitations~\cite{henn_emergence_2009,neely_characteristics_2013,navon_emergence_2016,navon_synthetic_2019}. Spectral analysis forms a central tool for understanding quantum fluid turbulence~\cite{billam_onsager-kraichnan_2014,simula_emergence_2014,billam_spectral_2015,chesler_holographic_2013,horng_two-dimensional_2009,karl_strongly_2017,johnstone_evolution_2019,mikheev_low-energy_2019,kobyakov_turbulence_2014,koniakhin_2d_2020,navon_emergence_2016,navon_synthetic_2019, numasato_direct_2010,reeves_inverse_2013,reeves_signatures_2014,reeves_enstrophy_2017,skaugen_vortex_2016,tsubota_quantum_2013,muller_abrupt_2020,nore_kolmogorov_1997,shukla_turbulence_2013}, providing a compact representation of kinetic features at widely different scales $\ell$ in terms of wavenumber $k=2\pi/\ell$; for reviews see Refs.~\cite{barenghi_introduction_2014,tsubota_turbulence_2014,white_vortices_2014,madeira_quantum_2020-1}. Classical spectral analysis was extended to compressible quantum fluids by introducing a generalized velocity field that uses the spatially-dependent density to remove the velocity singularity encountered at a vortex core~\cite{nore_kolmogorov_1997}, while retaining the formulation of kinetic energy as a quadratic form in the generalized velocity field. 

The identification of power-law behavior signifying turbulent cascades requires an accurate representation of kinetic energy in $k$-space. Famously, the Kolmogorov $-5/3$ power law~\cite{kolmogorov_local_1941} signifies an inertial energy cascade through scale space. First observed in ocean currents~\cite{grant_turbulence_1962}, numerical evidence for the $\sim k^{-5/3}$ energy spectrum has been reported in fully-developed turbulent flows in a variety of quantum fluid settings including the two-dimensional (2D) point-vortex model~\cite{novikov_dynamics_1975,skaugen_origin_2017}, and in 3D and 2D Gross-Pitaevskii turbulence~\cite{nore_kolmogorov_1997,bradley_energy_2012}. However, the velocity power spectrum measures of classical turbulence have limitations for describing compressible quantum fluids due to their neglect of quantum phase information. Formally, the compressible and incompressible power spectra are not locally additive in $k$-space~\cite{reeves_signatures_2014}, precluding a strict interpretation in terms of energy fluxes~\cite{muller_abrupt_2020}. 
While some progress has been made by assuming decoupling of the compressible and incompressible energies~\cite{numasato_direct_2010}, a fully quantum spectral analysis is required to move forward.

The paper is structured as follows. In \sref{sec:background} we give a short introduction to the GPE, and the representation of the GPE energy in momentum space. In \sref{app:wkt} we develop a general mapping between power spectra and correlations suitable for compressible quantum fluids. In \sref{sec:compfluid} we apply this approach to the GPE, developing a number of useful formal properties of energy spectral densities and velocity power spectra. In \sref{sec:apps} we present applications to a 2D tapped ground state, and the ground state imprinted with quantum vortices. In \sref{sec:conc} we offer our concluding remarks and outlook.

\section{Background}\label{sec:background}
\subsection{Gross-Pitaevskii Equation}\label{sec:gpe}
Bose-Einstein condensation of trapped gases occurs at a critical temperature $T_c$ where the excited states are saturated and, upon further cooling, atoms are forced \emph{en masse} into the trap ground state. Well below the critical temperature, $T\ll T_c$, a nearly pure BEC with wavefunction (order parameter) $\psi$ has Gross-Pitaevskii energy 
\begin{align}\label{Hgp}
H&\equiv\int d^3\rr\; \frac{\hbar^2}{2m}|\nabla\psi|^2+V|\psi|^2+\frac{g}{2}|\psi|^4,
\end{align}
where $g=4\pi\hbar^2 a/m$ is the two-body interaction parameter for atoms with s-wave scattering length $a$, and $V(\rr,t)$ is the external trapping potential. In what follows we will suppress space-time arguments unless there is risk of ambiguity. There is a clear interpretation of each term, with
\begin{align}\label{ek}
E_\text{kin}&\equiv \frac{\hbar^2}{2m}\intr |\nabla\psi|^2,\\
\label{ev}
E_\text{pot}&\equiv \intr |\psi|^2V(\rr,t),\\
\label{ei}
E_\text{int}&\equiv \frac{g}{2}\intr |\psi|^4,
\end{align}
giving the total kinetic, potential, and interaction energies respectively.
The GPE may be written as a functional derivative
\ali{\label{gpe}
i\hbar\frac{\del\psi(\rr,t)}{\del t}&=\frac{\delta H}{\delta \psi^*(\rr,t)}= L\psi,
}
where we define the \emph{nonlinear GP-operator}
\ali{\label{Lgp}
L\psi&\equiv\left(-\frac{\hbar^2}{2m}\nabla^2+V+g|\psi|^2\right)\psi.
}
The GPE conserves the energy $H$ and total particle number
\ali{\label{Ngp}
N&=\intr |\psi|^2.
}
Despite its relative simplicity, the GPE exhibits an extremely rich phenomenology.
\subsection{Madelung transformation}
Before exploring some of the essential physical phenomena of the GPE, it is helpful to define some further properties. 

The order parameter can be written as
\begin{align}\label{mad}
\psi&=\sqrt{n(\rr,t)}e^{i\Theta(\rr,t)}
\end{align}
in terms of the \emph{particle density} $n(\rr,t)=|\psi(\rr,t)|^2$, and the \emph{phase} $\Theta(\rr,t)$. In these hydrodynamic variables we have the superfluid velocity  
\ali{\label{velocity}
\mathbf{v}(\rr,t)&\equiv\frac{\hbar}{m}\nabla\Theta(\rr,t)}
and current density
\ali{\label{current}
\mathbf{J}(\rr,t)&\equiv \frac{i\hbar}{2m}\left(\psi\nabla\psi^*-\psi^*\nabla\psi\right)=n \mathbf{v},
}
and the Hamiltonian becomes
\ali{\label{hydroH}
H=\intr \frac{m}{2}n|\vv|^2+\frac{\hbar^2}{2m}
|\nabla\sqrt{n}|^2+Vn+\frac{gn^2}{2}.
}
The kinetic energy has been written in terms of a hydrodynamic term and a quantum pressure term. The hydrodynamic term is the primary focus of quantum turbulence studies as it allows identification of links with classical turbulence in \emph{incompressible} fluids. In a BEC this regime occurs when there is a static, homogeneous background density $n(\rr,t)\approx \bar n$, usually associated with a high chemical potential and hard-wall confinement~\cite{gaunt_bose-einstein_2013,gauthier_direct_2016}, with ground state wavefunction $\psi_0(\rr)$, satisfying $\mu\psi_0\equiv L\psi_0$. A planar BEC containing vortices is a special case of this regime, where the vortices move in a static background density~\cite{gauthier_giant_2019}; when  the cores are well separated the system can enter an approximate point-vortex regime. However, being compressible, BECs also exhibit significant coupling between Bogoliubov excitations and vortices, and analysis of kinetic energy requires decomposition into compressible (acoustic) and incompressible (vortex) parts~\cite{nore_kolmogorov_1997,bradley_energy_2012}.

\subsection{Compressible-incompressible decomposition}\label{sec:1C}
The hydrodynamic kinetic energy can be 
expressed as a quadratic form in the field 
\begin{align}\label{udef}
    \uu(\rr)\equiv \sqrt{n}(\rr)\vv(\rr)
\end{align}
This choice of density weighting regularizes the singular behavior near a vortex core arising from the $1/r$ divergence of the velocity field ($\sqrt{n}\sim r$ near the core). The expression of energy as a quadratic form in $\uu$ allows application of Parseval's theorem to derive power spectra.

For finite systems, the definition \eref{udef} also allows a Helmholtz decomposition~\cite{nore_kolmogorov_1997},  $\uu(\rr)= \uu^i(\rr)+\uu^c(\rr)$, with  incompressible and compressible components satisfying
\begin{align}\label{uidef}
    \nabla\cdot\uu^i(\rr)&=0, \\
    \label{ucdef}
    \nabla\times\uu^c(\rr)&=\mathbf{0},
\end{align}
 respectively~\footnote{The usual conditions on the decomposition hold. In three dimensions, it is equivalent to the existence of a vector potential $\mathbf{A}$ and a scalar potential $\phi$, such that $\mathbf{u}=\nabla\phi+\nabla\times\mathbf{A}$. Furthermore, the field should be twice continuously differentiable, and $|\mathbf{u}|$ should vanish faster than $1/r$ as $r\to\infty$}. The decomposition is easily effected in momentum space, see Appendix \ref{app:helm}. The vectors fields are orthogonal in momentum space by construction~\footnote{Note that there exist potential solutions, e.g. the quantum phase $\Theta(\rr)\propto x^2-y^2$, that will generate velocity fields that satisfy both \eref{uidef}, \eref{ucdef}. Such harmonic potential flows also arise in uniform translation or rotation and can be handled by subtracting the harmonic part of the velocity field explicitly before carrying out Helmholtz decomposition.}. 
Defining the quantum pressure velocity field 
\begin{align}\label{qpv}
    \uu^q &\equiv \frac{\hbar}{m}\nabla\sqrt{n},
\end{align}
the total kinetic energy can be written as
\begin{align}\label{Ekin:each}
    E_\text{kin}&=E_\text{kin}^i+E_\text{kin}^c+E_\text{kin}^q,
\end{align} 
where 
\begin{align}\label{Ekicq}
    E_\text{kin}^\alpha&=\frac{m}{2}\int d^d\rr\; |\uu^\alpha(\rr)|^2,
\end{align}
for $\alpha \in \{i,c,q\}$ \footnote{Note that the quantum pressure is strictly a compressible effect. However, since it does not depend on explicitly on the fluid velocity, we separate it from the compressible part of the hydrodynamic kinetic energy.}. Note that due to \eeref{qpv}, we have  $\nabla\times\uu^q=\mathbf{0}$, and the quantum pressure is formally compressible. We emphasize here that while we use the $\uu^\alpha$ notation to represent all components, only $\uu^c$ and $\uu^q$ are physical fluid velocities related to $\vv$ itself, while $\uu^q$ merely has the physical dimension of a velocity.
\subsection{Velocity power spectra and energy spectral densities}
It is worthwhile to briefly state some useful results from Fourier analysis. 
Parseval's theorem allows any integral expressed as a quadratic form of the kind \eref{Ekicq} to be written as an integral in a corresponding frequency domain:
\begin{align}\label{ehpars}
    E_\text{kin}^\alpha&=\frac{m}{2}\int d^d\kk\;|\tilde{\uu}^\alpha(\kk)|^2,
\end{align}
where 
\begin{align}
    \tilde{\uu}^\alpha(\kk)&\equiv \frac{1}{(2\pi)^{d/2}}\int d^d\rr\;e^{-i\kk\cdot\rr}\uu^\alpha(\rr),
\end{align}
and we use the usual shorthand $\uu^\alpha(\rr)\longleftrightarrow \tilde{\uu}^\alpha(\kk)$.
As a concrete example, in 2D, we can move to cylindrical coordinates in $k$-space to find
\begin{align}
    E_\text{kin}^\alpha&=\int_0^\infty dk\;\frac{mk}{2}\int_0^{2\pi} d\theta_k \;|\tilde{\uu}^\alpha(k\cos\theta_k,k\sin\theta_k)|^2\notag\\\label{ecspecdens}
    &\equiv \int_0^\infty dk\;\varepsilon_\text{kin}^\alpha(k),
\end{align}
with the latter serving to define the velocity power spectral density, or simply the ``power spectrum". We have taken care not to refer to this as a spectral density. In a compressible superfluid we can identify three sources of kinetic energy, \eref{Ekin:each}. While each form of energy may be written as an integral similar to \eeref{ecspecdens} using Parseval's theorem, summing to the total kinetic energy, it is well understood in Fourier analysis that the integrands do not sum to the energy density locally at $k$. To be explicit, in suitable coordinates we can always write the total kinetic energy as
\begin{align}\label{Ektot}
    E_\text{kin}&=\frac{\hbar^2}{2m}\int d^d\rr\; \left|\nabla\psi\right|^2\equiv \int_0^\infty dk\;e_\text{kin}(k),
\end{align}
where the integrand serves to define the spectral density. For example, in 2D we have
\begin{align}\label{Ekdens2}
    e_\text{kin}(k)&=\frac{\hbar^2 k}{2m}\int_0^{2\pi}d\theta_k\;\left|\nabla\psi\right|^2,
\end{align}
where $e_\text{kin}(k)dk$ is the amount of kinetic energy in the range $[k,k+dk)$.
However, the local additive property does not hold for the velocity power spectra obtained using Parseval's theorem: 
\begin{align}\label{nonadd}
    e_\text{kin}(k)&\neq \varepsilon_\text{kin}^i(k)+\varepsilon_\text{kin}^c(k)+\varepsilon_\text{kin}^q(k).
\end{align}
 One of our main aims of this paper is to reconcile the decomposition \eref{Ekin:each} with the spectral representation \eref{ecspecdens}. A constructive statement of the true spectral densities for a compressible quantum fluid of the Gross-Pitaevskii type will be given in \sref{kedtot}. Note also that the mapping to Fourier space simplifies for an incompressible classical fluid: only the incompressible kinetic energy remains and hence the additivity property is not required, and Parseval's theorem gives $E_\text{kin}\to E_\text{kin}^i$, and $\varepsilon_\text{kin}^i(k)$ is a legitimate spectral density.

We can clarify the situation by noting that the standard spectral measures applied to turbulent BECs~\cite{bradley_energy_2012,horng_two-dimensional_2009,koniakhin_2d_2020,neely_characteristics_2013,nore_kolmogorov_1997} in the form \eref{ecspecdens} are not energy spectral densities, but rather power spectral densities of autocorrelation functions:
\begin{align}\label{2pc}
\int d^d\rr\;\uu^\alpha(\rr)^*\cdot \uu^\alpha(\rr+\rr')&\longleftrightarrow  (2\pi)^{d/2}|\tilde{\uu}^\alpha(\kk)|^2.
\end{align}  
The quantity on the left is the two-point correlator for the vector field, while its Fourier transform represents the same information in scale space. Such quantities provide a wealth of useful semi-classical information about superfluid dynamics, and are often loosely referred to as ``energy spectra". 

Numerous works have used this Parseval theorem approach to compute velocity power spectra for GPE simulations of atomic~\cite{nore_kolmogorov_1997,araki_energy_2002,kobayashi_kolmogorov_2005,parker_emergence_2005,horng_two-dimensional_2009,bradley_energy_2012,reeves_signatures_2014} and polariton BECs~\cite{koniakhin_2d_2020}. As the simulations are almost always performed on cartesian grids, the spectra are typically computed by binning the data in $k$ space. Binning involves computing a numerical approximation to the angular integral by summing over annuli $k_1 < k < k_2$. For cartesian data this process is inherently a rather crude approximation, particularly in the small-$k$ regime where the point density in each bin becomes sparse. This motivates the approach developed in the present work, which instead evaluates the $k$-space angular integrals analytically. A further benefit of this approach is a decoupling of the grid resolution of the wavefunction, from the $k$-space resolution of the resulting spectra. The latter is entirely flexible, allowing high resolution spectra to be easily computed. This also allows for accurate construction of angle-averaged two-point correlations by Fourier inverting the spectra to position space.

\section{Compressible quantum fluids}\label{sec:compfluid}
Our aim is to construct accurate spectral densities in momentum space using spatial data for the wavefunction~$\psi(\rr,t)$. Typically numerical simulations of the GPE are computed on cartesian grids, while the spectral density used for analysis is a function of $k$. We approach this problem by formally writing a general vector inner product as a spectral density, carrying out the angular integrals analytically. As a corollary we develop an angle-averaged Wiener-Khinchin theorem relating a spectral density to an associated correlation function. We focus on dimensions 2 and 3, as in 1D the problem reduces to mapping a two-sided power spectrum to a one-sided power spectrum.

A complete description of the proof is given in Appendix \ref{app:wkt}. Here we summarize the main results. 
Given two complex valued vector fields $\uu$, $\vv$, we define the spectral density of their inner product, $\langle\uu||\vv\rangle(k)$,  as 
\begin{align}\label{specdens_main}
    \langle \uu|\vv\rangle&\equiv\int_0^\infty dk\;\langle\uu||\vv\rangle(k),
\end{align}
and this is shown to be given by 
\begin{align}\label{specdens_main}
    \langle\uu||\vv\rangle(k)&\equiv\int d^d\x\;\Lambda_d(k,|\x|)C[\uu,\vv](\x),
\end{align}
where the two-point correlation in position is
\begin{align}\label{Cdef_main}
    C[\uu,\vv](\x)&\equiv\int d^d\mathbf{R}\;\ip {\uu}{\mathbf{R}-\x/2}\ip{\mathbf{R}+\x/2}{\vv}, 
\end{align}
and the kernel function is dimension dependent:
\begin{align}\label{Lamdef_main}
    \Lambda_d(k,r)&\equiv
    \begin{cases}
        \frac{1}{2\pi}\;kJ_0(kr),&\text{for } d=2,\\
        \frac{1}{2\pi^2}\;k^2\sinc{(kr)},&\text{for } d=3.
    \end{cases}
\end{align}
The cartesian two-point correlation is thus directly mapped to a spectral density function of $k$ by formal integrating over angles in $k$-space. A further useful consequence of this formulation is that we also find an exact mapping from $k$-space back to an angle-average of the two-point correlation function \eeref{Cdef_main}, which takes the form
\begin{align}\label{guv_main}
    g_{uv}(r)\equiv\int_0^\infty dk\;\Lambda_d^{-1}(k,r)\frac{\langle\uu||\vv\rangle(k)}{\langle\uu|\vv\rangle },
\end{align}
with inverse kernel
\begin{align}\label{Laminvdef_main}
    \Lambda_d^{-1}(k,r)&\equiv\begin{cases}
        J_0(kr)&\text{for } d=2,\\
        \sinc{(kr)}&\text{for } d=3.
    \end{cases}
\end{align}
We have absorbed the value of $C[\uu,\vv](\mathbf{0})=\langle\uu |\vv\rangle$ for convenience, so that $g_{uv}(0)\equiv 1$ by definition. We can summarize this as an angle-averaged Wiener-Khinchin theorem:
\begin{align}\label{wksum_main}
    (\uu,\vv)\longrightarrow \langle \uu||\vv\rangle(k)\longleftrightarrow g_{uv}(r),
\end{align}
where the first mapping is a Fourier transform followed by angular integration in $k$ space. The second is the Fourier relation between the $k$ and $r$ variables. Our analysis amounts to an explicitly angle-averaged formulation of the standard Wiener-Khinchin theorem linking power spectra with two-point correlation functions. Our development in terms of general vector fields allows many quantities of interest to be calculated for compressible quantum fluids, by choosing $\uu$ and $\vv$ appropriately. We will consider examples below.

Explicit angular integration offers formal and numerical advantages for spectral analysis, particularly in the context of quantum turbulence~\cite{nore_kolmogorov_1997,bradley_energy_2012}. The formal gains will appear clearly in the following sections where we apply these results to energy spectral densities, enabling a rigorous decomposition of the kinetic spectral density into compressible, incompressible, and quantum pressure terms, and their mutual interactions. 

The numerical advantages are twofold. First, we can exploit standard Fourier methods to efficiently evaluate the convolution, \eref{Cdef_main}. Second, by writing the power spectrum \eref{specdens_main} and the two-point correlation \eref{guv_main} in terms of the kernels \eref{Lamdef_main} and \eref{Laminvdef_main}, we have decoupled the grid of the vector fields from the grid of the spectrum and correlator: the kernel functions can be evaluated on \emph{any} desired $k$ or $r$ grids to create the desired spectra or correlator. This avoids the need to bin cartesian grid data or interpolate onto a suitable grid for angular integration; it also allows for rectangular spatial domains, as the convolution may be evaluated for vector fields with arbitrary coordinate range.
Moreover, for computing spectra spanning many orders of magnitude in turbulent flows it is often desirable to have a linear point density in log space --- a choice that can be freely made in the present formulation. 

We now apply the angle-averaged Wiener-Khinchin theorem derived in Appendix \ref{app:wkt} to central problems of spectral analysis in compressible quantum fluids. 

\subsection{Velocity Power Spectrum} 
We return now to the compressible, incompressible, and quantum pressure kinetic energies \eref{ehpars}.
Taking $\mathbf{u}=\mathbf{v}\equiv \mathbf{u}^\alpha$ in \eref{wksum_main}, we have

\begin{align}\label{epsint}
    E_\text{kin}^\alpha&=\int_0^\infty dk\;\varepsilon_\text{kin}^\alpha(k),
\end{align}
with velocity power spectrum $\varepsilon_\text{kin}^\alpha(k)\equiv\frac{m}{2}\langle \mathbf{u}^\alpha||\mathbf{u}^\alpha\rangle(k)$ given by
\begin{align}\label{Ekkic}
    \varepsilon_\text{kin}^\alpha(k)&=\frac{m}{2}\int d^d\x\;\Lambda_d(k,|\x|)C[\uu^\alpha,\uu^\alpha](\x).
\end{align}
We have arrived at a formulation of the velocity power spectrum that may be quickly and accurately evaluated on a cartesian grid using a discrete Fourier transform. 
The interpretation is also straightforward: for any of the position-space fields $\uu^\alpha$, there is a spectral density, \eref{Ekkic}, that is equivalent to an angle-averaged two-point correlation \eref{2pc} represented in $k$ space. 

\subsection{Kinetic Spectral Density}\label{kedtot}

In this section we derive a spectral decomposition of the kinetic energy that does not rely on Parseval's theorem. Unlike the Parseval approach, the resulting quantities are true spectral densities that are locally additive in $k$ space. They include all quantum phase information, and provide a natural starting point for computing energy fluxes and other measures of quantum turbulence. 

For the total kinetic energy \eref{Ektot}, we find the spectral density in the form
\begin{align}\label{ekintot}
    e_\text{kin}(k)&\equiv \frac{\hbar^2}{2m}\langle \nabla\psi||\nabla\psi\rangle(k),
\end{align}
equivalent to the integrated two-point correlation function 
\begin{align}\label{Ektot2}
    e_\text{kin}(k)&=\frac{\hbar^2}{2m}\int d^d\x\;\Lambda_d(k,|\x|)C[\nabla\psi,\nabla\psi](\x),
\end{align}
used for fast numerical evaluation.
The total kinetic energy \eref{Ektot} is immediately recovered by integrating \eref{Ektot2} over all $k$ and using \eref{deltagen}. We note that this is also equivalent to the Fourier representation
\begin{align}
    E_\text{kin}&=\frac{\hbar^2}{2m}\int d^d\kk\;|\kk\phi(\kk)|^2,
\end{align}
where $\phi(\kk)$ is the Fourier transform of $\psi(\rr)$. In momentum space we can thus identify the kinetic spectral density as
\begin{align}\label{ekindef}
    e_\text{kin}(k)&=\frac{\hbar^2k}{2m}\int d\Omega_d|\kk\phi(\kk)|^2,
\end{align}
where we integrate over the $k$-space solid angle in $d$ dimensions. However, the decomposition into compressible, incompressible and quantum pressure contributions is more direct in position space; in what follows we primarily use \eeref{Ektot2} for numerical evaluation.

\subsection{Decomposed Kinetic Energy Density}\label{decompE}
With \eeref{Ektot2} in hand, we can now decompose the total kinetic energy density into the kinetic energy density of the compressible and incompressible parts, and the quantum pressure.

We retain all phase information and apply a Madelung transformation \eref{mad}, and a Helmholtz decomposition. Evaluating the gradient of $\psi$ in terms of $n$ and $\Theta$, we find 
\begin{align}\label{psigrad}
    \nabla\psi&=\frac{m}{\hbar}[\ww^i+\ww^c+\ww^q],
\end{align}
where the imaginary unit and phase factors are absorbed into the vector fields
\begin{align} \label{wdef}
    (\ww^i,\ww^c,\ww^q)&\equiv(i\uu^i,i\uu^c,\uu^q)e^{i\Theta}.
\end{align}
Equation \eref{psigrad} contains all information about kinetic energy in a quantum fluid. We use it here to promote the well-known semiclassical spectral analysis to a quantum treatment that includes all phase information. Before we proceed, it important to note that these fields do not inherit the properties of $\uu^\alpha$: due to the quantum phase, we have in general $\nabla \times \ww^c\neq\mathbf{0}$, $\nabla\cdot\ww^i\neq 0$, and $\nabla \times \ww^q\neq\mathbf{0}$; consequently, the following decomposition makes no assumptions about compressibility or otherwise of the $\ww$ fields, relying only on properties of $\uu^\alpha$.

Proceeding with our analysis of \eref{ekintot} using the decomposition \eref{psigrad}, we can write the vector convolution in \eref{Ektot2} as 
\begin{align}
    \frac{\hbar^2}{m^2}C[\nabla\psi,\nabla\psi](\x)&=\sum_{\alpha,\beta}C[\ww^\alpha,\ww^\beta](\x),
\end{align}
and the complete kinetic spectral density may now be written as
\begin{align}\label{ekindens}
    e_\text{kin}(k)&\equiv e_\text{kin}^i(k)+e_\text{kin}^c(k)+e_\text{kin}^q(k)\notag\\
    &+e_\text{kin}^{ic}(k)+e_\text{kin}^{iq}(k)+e_\text{kin}^{cq}(k),
\end{align}
where 
\begin{align}\label{Ekdens}
    e_\text{kin}^\alpha(k)&\equiv\frac{m}{2}\int d^d\x\;\Lambda_d(k,|\x|)C[\ww^\alpha,\ww^\alpha](\x),\\
    e_\text{kin}^{\alpha\beta}(k)&\equiv\frac{m}{2}\int d^d\x\;\Lambda_d(k,|\x|)\Big\{C[\ww^\alpha,\ww^\beta](\x)\nonumber\\
    &+C[\ww^\beta,\ww^\alpha](\x)\Big\}.\label{eijdense}
\end{align}
In this form we use the angle-averaged Wiener-Khinchin theorem of Appendix \ref{app:wkt} to establish a number of important properties of the kinetic spectral density.

Using \eeref{deltagen}, the integral of each spectral density gives (just as we found for the velocity power spectrum, \eeref{epsint})
\begin{align}\label{Ekint}
    \int_0^\infty dk\;e_\text{kin}^\alpha(k)&=\frac{m}{2}\int d^d\rr\;|\uu^\alpha(\rr)|^2=E_\text{kin}^\alpha,
\end{align}
since the exponential phase factors in $C[\ww^\alpha,\ww^\beta](\x)$ cancel at $\x=0$.  
Moreover, as shown in Appendix \ref{app:coupling}, the coupling terms all integrate to zero:
    \begin{align} 
    \int_0^\infty dk\;e_\text{kin}^{\alpha\beta}(k)&=0,\quad\quad \alpha\neq\beta.\label{Eabint}
\end{align}

We immediately recover the total kinetic energy by integrating over \eeref{ekindens}: 
\begin{align}
    E_\text{kin}&=\int_0^\infty dk\;[e_\text{kin}^i(k)+e_\text{kin}^c(k)+e_\text{kin}^q(k)].
\end{align}

This shows two important features of the quantum kinetic energy density. First, $e_\text{kin}^{\alpha\beta}(k)$ does not contribute to the total energy, but instead redistributes energy between different scales. These terms hence describe an interaction between different forms of energy. Second, the quantum phase information present in \eref{wdef} is essential to obtain locally additive spectral densities in $k$ space~\cite{reeves_signatures_2014}, required for interpreting changes in spectra as energy transport phenomena. Neglecting the quantum phase in \eref{wdef} imposes a semiclassical approximation. The formal replacement $e^{i\Theta}\to 1$ is equivalent replacing quantum spectral densities with semiclassical velocity power spectra: $e_\text{kin}^\alpha(k)\to\varepsilon_\text{kin}^\alpha(k)$. The velocity power spectrum measures of turbulence thus focus on a reduced set of degrees of freedom associated with classical turbulence. As such they are useful for understanding links between quantum and classical fluids, as explored extensively in previous works~\cite{kobayashi_quantum_2021,nore_kolmogorov_1997,tsubota_quantum_2013}

A spectral density can also be found for the potential and interaction energy terms using \eref{specdens} to correlate the fields $\sqrt{nV}$ and $n$ respectively:
\begin{align}
    e_\text{pot}(k)&=\int d^d\x\;\Lambda_d(k,|\x|)C[\sqrt{nV},\sqrt{nV}](\x),\\
    e_\text{int}(k)&=\frac{g}{2}\int d^d\x\;\Lambda_d(k,|\x|)C[n,n](\x).
\end{align}
Such quantities are automatically locally additive, as the quantum phase factor plays no role in the corresponding term in~\eref{Hgp}. Hence, the total system energy can be decomposed into locally additive spectral densities in $k$ space as
\begin{align}
    e_\text{tot}(k)&=e_\text{kin}(k)+e_\text{pot}(k)+e_\text{int}(k),
\end{align}
where $e_\text{kin}(k)$, defined in \eeref{ekindef}, may be further decomposed using \eref{ekindens}, and the coupling terms $e_\text{kin}^{\alpha\beta}(k)$ are in general non-vanishing.

\section{Applications}\label{sec:apps}
We consider three readily accessible systems in which to apply the spectral analysis, in two dimensions: a pure GPE ground state in a harmonic trap, a GPE ground state with a single central vortex, and a GPE ground state with a neutral vortex distribution. We compute spectral densities and correlation functions. For the first two systems, we also compare with analytical results.  

\subsection{Thomas-Fermi state in a 2D Harmonic trap}\label{sec:thomasfermi}
As a first application we compute a range of spectral densities for the ground state of the GPE in an oblate cylindrical harmonic trap
\begin{align}
    V(\rr)&=\frac{m}{2}(\omega_r^2(x^2+y^2)+\omega_z^2z^2),
\end{align}
where $\omega_z\gg\omega_r$. As the ground state contains no velocity field, all velocity power spectra are zero.
Integrating over a tightly confined $z$ dimension, assumed in the ground state 
\begin{align}
    \phi_{0}(z)&=\left(\frac{1}{\pi a_z^{2}}\right)^{1 / 4} e^{-z^{2} / 2 a_z^{2}}
\end{align}
with oscillator length $a_z=\sqrt{\hbar/m\omega_z}$,
the effective interaction strength becomes 
\begin{align}
    g_2&=\frac{g}{\sqrt{2\pi a_z^2}}.
\end{align}
For $N$ atoms in such a trap, we use the Thomas-Fermi (TF) wavefunction
\begin{align}\label{tfwave}
    \psi_\text{TF}(\rr)&=\begin{cases}
        \sqrt{n_0}\sqrt{1-r^2/R^2},\quad&\text{for}\;r\leq R,\\
        0,\quad&\text{for}\;r>R.
        \end{cases}
\end{align}
for $r=\sqrt{x^2+y^2}$, peak density $n_0=\mu/g_2$, and TF-radius $R=\sqrt{2\mu/m\omega_r^2}$. The Thomas-Fermi chemical potential is then
\begin{align}
    \mu&=\hbar\omega_r\left(\sqrt{\frac{8}{\pi}}\frac{aN}{a_z}\right)^{1/2}.
\end{align}
We consider a system of $N=1.21\times 10^5$ $^{87}$Rb atoms in a trap with angular frequencies $\omega_z=2\pi\times 100$ Hz, $\omega_r=2\pi\times 3$ Hz. In units of $\hbar\omega_r$ and $a_r=\sqrt{\hbar/m\omega_r}$, we have $\mu=30$, $g_2=0.0233\hbar\omega_r a_r^2$, and the Thomas-Fermi radius and healing length are $R=7.46a_r$, and $\xi = \hbar/\sqrt{m\mu}=0.18a_r$ respectively. The system is thus in the Thomas-Fermi regime $\xi\ll R$. In all of our computations, we use a spatial domain of side length $22a_r$, with $512$ points, so that $\Delta x=\Delta y = 0.043a_r$. As our reference state is a Thomas-Fermi ground state, we present our results in units of $\xi$ and $\mu$, and in these units $a_r = 5.6\xi$, $g_2 = 0.0239\mu\xi^2$, and $R = 41.4\xi$.

The spectral densities can be calculated analytically for the kinetic, trap, and interaction energies.
We start with the kinetic term. First we note that as the ground state has no phase gradient, the kinetic energy is entirely due to $\nabla\sqrt{n}$; we can hence compare the analytical kinetic spectral density with the numerical quantum pressure spectral density. The analytical results are more easily computed directly in Fourier space, while numerically we use the expression in terms of the integrated spatial field correlation. Moving to polar coordinates and using cylindrical symmetry, we have
\begin{align}\label{ekin_all}
    e_\text{kin}(k)&=\frac{\hbar^2k^2}{2m}2\pi k|\phi(\kk)|^2.
\end{align}
We make the Thomas-Fermi approximation, neglecting the kinetic energy from the BEC boundary at $r\sim R$. The pure TF state has no velocity field contribution to the kinetic energy and the quantum pressure term is the only non-vanishing term. Proceeding as in Appendix \ref{app:tf}, we can write the kinetic, interaction, and potential energy densities in the form
\begin{align}\label{Etfkin}
    e_{\mathrm{kin,TF},a}(k)&=\bar\epsilon F(kR),\\
    \label{Etfint}
    e_{\mathrm{int,TF},a}(k)&= \left(\frac{\mu}{\hbar\omega_r}\right)^2\bar\epsilon G(kR),\\
    \label{Etfpot}
    e_{\mathrm{pot,TF},a}(k)&= \left(\frac{\mu}{\hbar\omega_r}\right)^2 \bar\epsilon  H(kR), 
\end{align}
where the energy density unit is
\begin{align}\label{epsbar}
\bar\epsilon&\equiv \hbar^2n_0\pi R/m,
\end{align}
and 
\begin{align}
    F(q)&\equiv (\sin{q}-q\cos{q})^2/q^3,\\
    G(q)&\equiv J_2(q)^2(2/q)^3,\\
    H(q)&\equiv  \pi^2\left[qJ_0(q/2)-2J_1(q/2)\right]^2J_1(q/2)^2/q^3.
\end{align}
\begin{figure}[!t]
    \centering
    \includegraphics[width=\columnwidth]{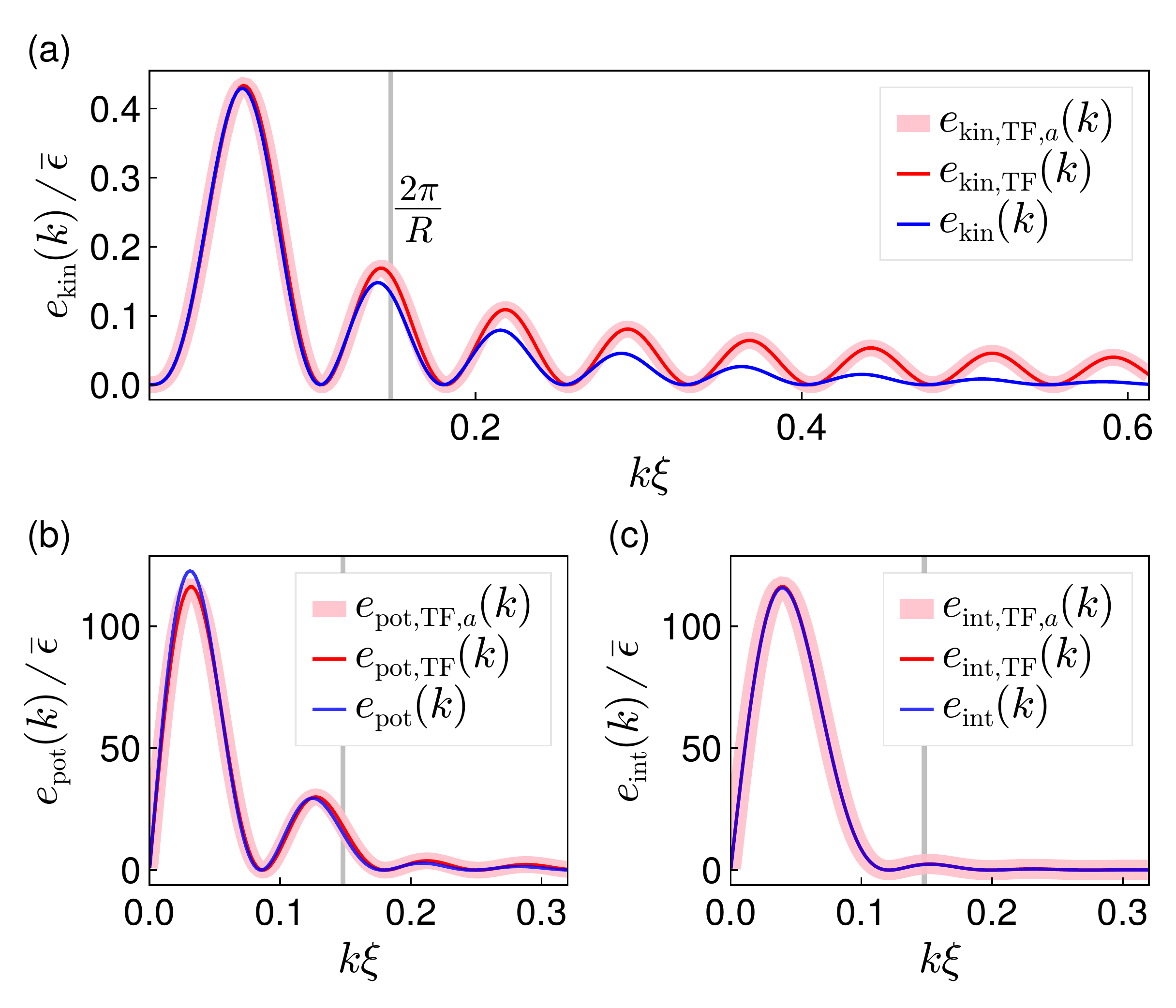}
    \caption{Energy densities for Thomas-Fermi and Gross-Pitaevskii wavefunctions in units of $\bar\epsilon$. (a) Total kinetic energy, (b) Potential energy, (c) interaction energy. The line at $k=2\pi/R$ shows the wavenumber associated with the TF radius $R$. }
    \label{fig1}
\end{figure}
In \fref{fig1} we compare kinetic, potential, and interaction energy densities for the Thomas-Fermi analytic results with numerical energy densities calculated for Thomas-Fermi and Gross-Pitaevskii solutions. In the TF treatment we should expect exact agreement between $e_\text{kin,a}(k)$ found by evaluating \eref{Etfkin}, and $e_\text{kin,TF}(k)$ computed numerically using the full spectral analysis on the TF wavefunction, in the form \eref{Ekdens}. Indeed we see excellent agreement between these two calculations, validating our general formulation of spectral analysis. We also see excellent agreement between the three approaches for $kR\ll 1$, as expected in the TF regime. Note that in general there is no restriction on the spectrum resolution in the infrared, in contrast to the binning approach that is strongly grid limited for small $k$. In the opposite regime, $kR\gg 1$, the sharp TF boundary generates a Bessel tail which dominates the spectra. The scale $k\sim 2\pi/R$ indicates the transition from the infrared features to the ultraviolet regime. The analytical result and the  numerical spectral analysis for the TF wavefunction are in excellent agreement for all $k$. 

The potential and interaction energy densities decay much more rapidly with $k$ than the kinetic energy density, strongly suppressing energy density beyond $k\sim 2\pi/R$. The interaction energy density shows excellent agreement for all $k$, while the potential energy agrees well except at the peak $k\xi\sim 0.05$. The difference arises from the GP solution having slightly lower density near $r\sim 0$ and higher density near $r\sim R$ than the TF state, a feature amplified by the radial trapping potential. 

\begin{figure}[!t]
    \centering
    \includegraphics[width=\columnwidth]{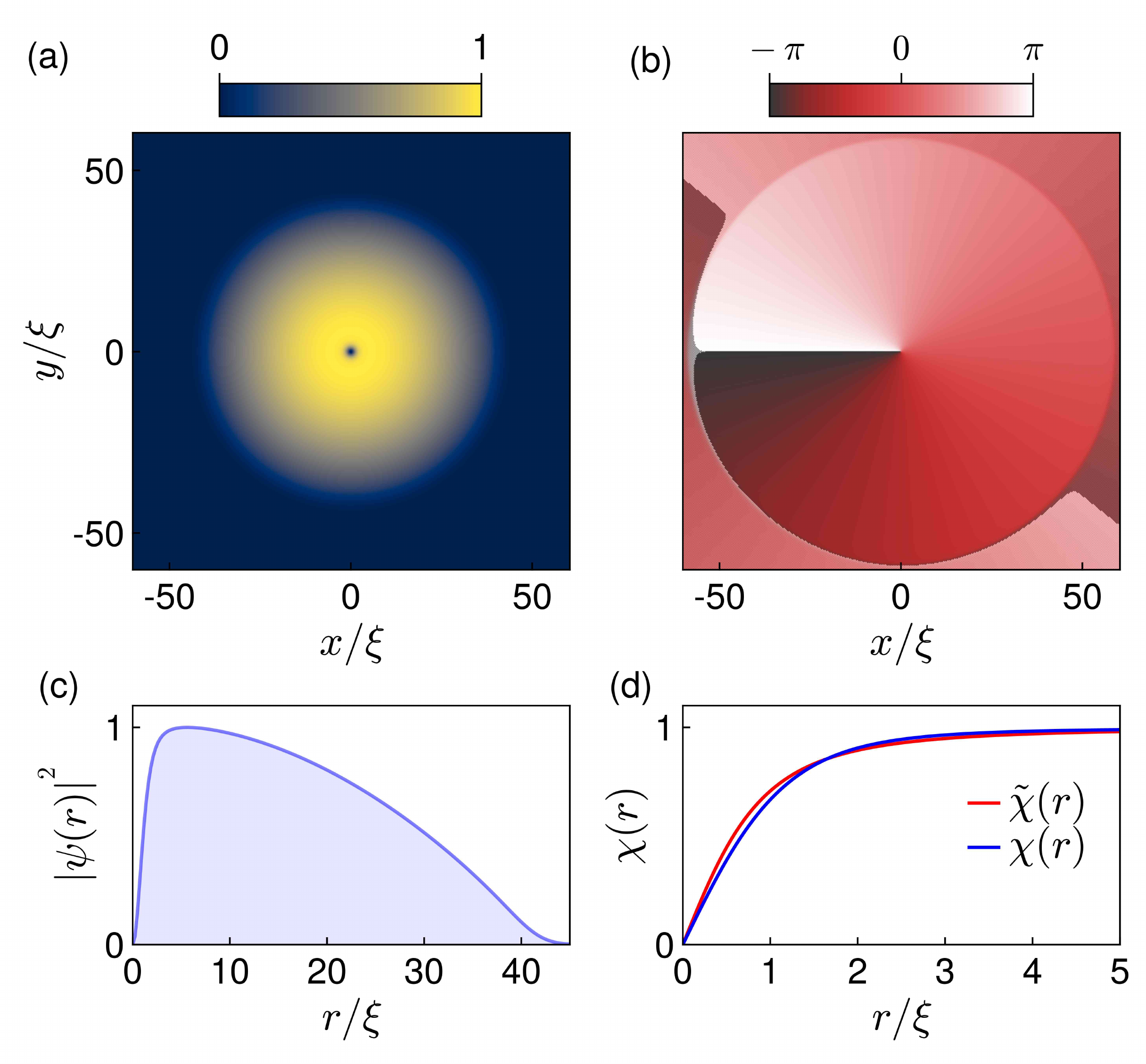}
    \caption{Gross-Piteaveskii ground state in a harmonic trap, imprinted with a charge-1 vortex: (a) atomic density $n(\rr)=|\psi(\rr)|^2$, relative to the maximum density; (b) quantum phase $\Delta=-(i/2)\log{(\psi/\psi^*)}$. The visible domain has side length $22a_r=120.5\xi$; (c) Radial slice through the atomic density, relative to the peak density; (d) vortex core ansatz $\tilde\chi(r)$, \eeref{psia}, and exact (numerical) vortex wavefunction $\chi(r)$ (see text).  }
    \label{fig2}
\end{figure}
\subsection{Central vortex: velocity power spectrum}
We now consider a trapped BEC containing a central vortex of charge 1. The state has a non-trivial velocity field, and interesting interactions between incompressible and compressible energies.  We use three different wavefunctions, as follows.
\begin{enumerate}[(a)]
    \item \emph{TF background, ansatz core.---} We construct a trapped state containing a central vortex by multiplying a smooth TF background wavefunction by an ansatz for the vortex core, with appropriate healing length for the local density, and the phase factor for a charge-1 vortex. This gives the Thomas-Fermi vortex (TF$v$) approximation  to the wavefunction
    \begin{align}\label{tfav}
        \psi_{\mathrm{TF}v,a}(\rr)&=\psi_\text{TF}(\rr)\tilde\chi(r)e^{i\theta},
    \end{align}
    where $\psi_\text{TF}(\rr)$ is given by \eeref{tfwave} and the vortex core is approximated as
    \begin{align}\label{psia}
        \tilde\chi(r)\equiv\frac{r}{\sqrt{r^2+(\xi/\Lambda)^2}},
    \end{align}
    and $\Lambda\simeq 0.8249$ ensures the correct slope at the vortex core~\cite{bradley_energy_2012}. The core function encodes the rapid variation of the density over the healing length $\xi=\hbar/\sqrt{m\mu}$, and $e^{i\theta}$ applies the quantum phase for a charge-1 vortex with winding $2\pi$ around the origin. This state gives an analytically tractable approximation to the true GPE vortex. 
    \item \emph{GP ground state background, GP core.---} The GPE ground state for the harmonic trap, found using imaginary time evolution, is imprinted with a numerically exact GP-vortex core
    \begin{align}\label{gpev}
        \psi_{v}(\rr)&\equiv\psi_\text{GP}(\rr)\chi(r)e^{i\theta},
    \end{align}
    where the core shape $\chi(r)$ is found numerically by solving the single vortex GPE~\cite{fetter_rotating_2009} on an infinite domain using a Chebychev basis. Since $\xi\ll R$, this approach gives a very accurate approximation to the GPE vortex ground state; it has also been used to construct initial states for dipole dynamics, with negligible compressible energy introduced by the imprinting~\cite{cawte_snells_2019}. The state $\psi_{v}(\rr)$ is shown in \fref{fig2} (a-c). In \fref{fig2}(d) we show the exact (numerical) vortex radial wavefunction $\chi(r)$, and the core ansatz wavefunction $\tilde\chi(r)$, \eeref{psia}. The ansatz has the same asymptotics for $r\ll \xi$ and $r\gg \xi$ as the GP vortex core, but differs slightly in the intermediate region $r\sim\xi$. 
\end{enumerate}

We calculate the velocity power spectrum numerically using \eeref{Ekkic}, and compare the result with two analytical approaches. 

First, a homogeneous approximation may be found by considering the vortex core of \eref{tfav} in a uniform BEC with density $n_0$~\cite{bradley_energy_2012}. The velocity power spectrum is
\begin{align}
    \varepsilon_{\text{kin},h}^i(k)= \pi n_0\xi^3\mu  F_\Lambda(k\xi)
\end{align}
where $F_\Lambda(q)=f(q/2\Lambda)^2/q$, and   
\begin{align}
    f(q)&=q(I_0(q)K_1(q)-I_1(q)K_0(q)),
\end{align}
and $K_\nu(q)$ and $I_\nu(q)$ are modified Bessel functions of the first and second kind, respectively. 
This velocity power spectrum has universal asymptotics in the IR and UV given by $F_\Lambda(q)\to 1/q$, and $F_\Lambda(q)\to \Lambda^2/q^3$ respectively. The former is the spectrum for an ideal point vortex, while the latter is the spectrum for a single compressible vortex solution of the Gross-Pitaevskii equation; the UV $k^{-3}$ regime is a well-known property of a single GP-vortex core stemming from the $\chi(r)\sim \Lambda r/\xi$ behavior as $r\to 0$~\cite{krstulovic_comment_2010-1,bradley_energy_2012}. 

\begin{figure}[!t]
    \centering
    \includegraphics[width=\columnwidth]{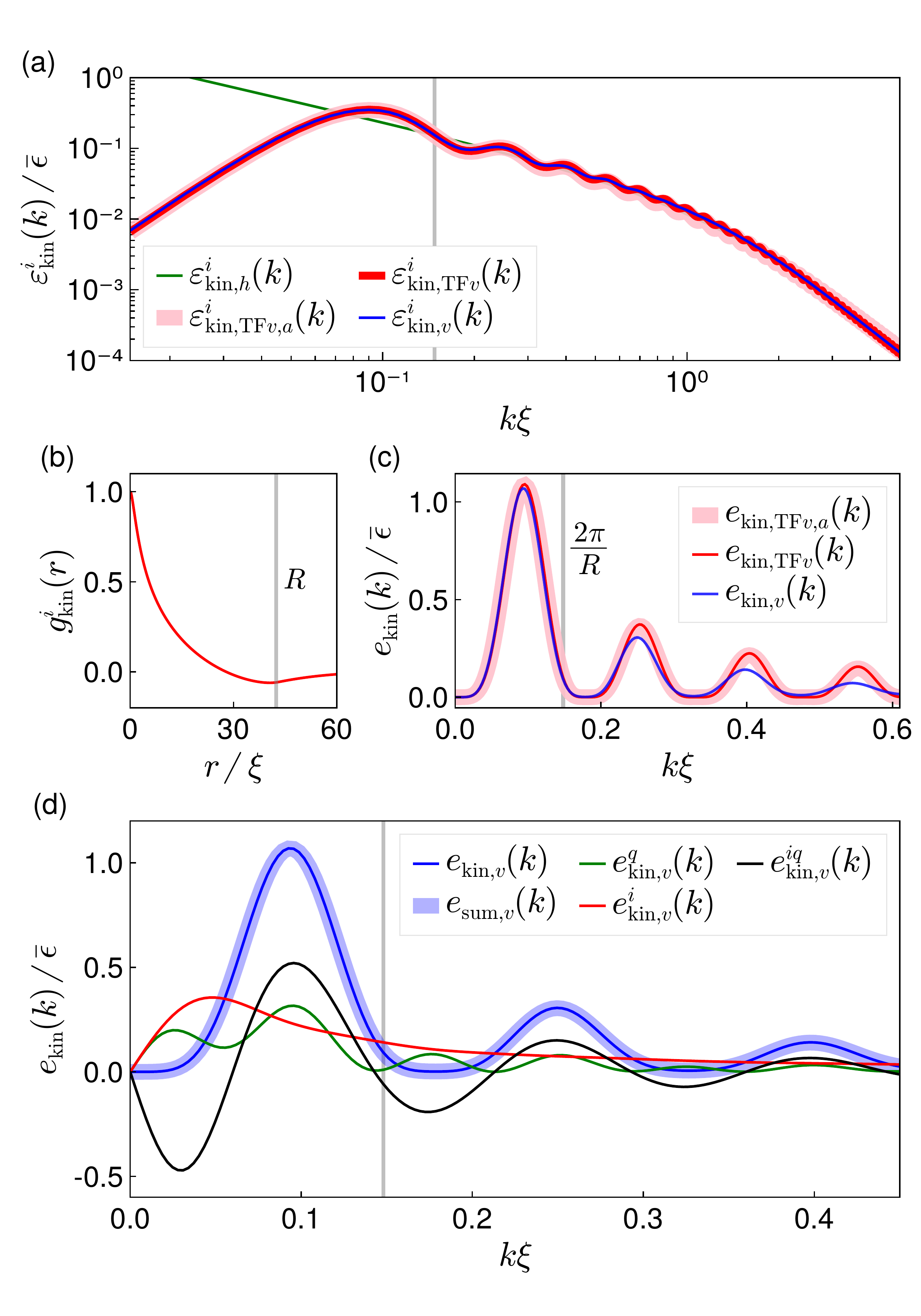}
    \caption{Spectral analysis of the central vortex in a trapped BEC: (a) Velocity power spectrum for the TF$v$ (thick red line), compared with the result of numerical spectral analysis of the GP vortex (thin red line). The analytic spectrum for a single vortex in a homogeneous BEC is shown for comparison~\cite{bradley_energy_2012} (thin blue). The two red curves agree for $k\xi\gg 1$ (within the vortex core). (b) Angle-averaged autocorrelation function of the incompressible velocity computed from $\varepsilon_\text{kin}^i(k)$ using \eref{Guv}, for the TF$v$ wavefunction. (c) Kinetic spectral density of the TF$v$, GPE, and analytic central vortex states. (d) Decomposed spectral densities of the GP vortex state, and coupling interaction, compared with the total kinetic spectral density for the GPE vortex state.}
    \label{fig3}
\end{figure}

Second, we can semi-analytically calculate the velocity power spectrum for $\psi_{\text{TF}v}(\rr)$. The density-weighted velocity field of the central vortex 
\begin{align}\label{uvr}
    \uu(\rr)&=\pm\frac{\hbar\sqrt{n_0}}{m}\sqrt{1-\frac{r^2}{R^2}}\frac{(-\sin\theta,\cos\theta)}{\sqrt{r^2+(\xi/\Lambda)^2}},
\end{align}
is entirely incompressible, $\uu(\rr)\equiv \uu^i(\rr)$, due to orthogonality between density gradient and the (divergence free) velocity field of a vortex.
As shown in \aref{app:tf}, the Fourier transform is 
\begin{align}\label{uvka}
    \tilde\uu^i(\kk)&=\mp i\frac{\hbar\sqrt{n_0}R}{m}(-\sin\theta_k,\cos\theta_k)T_1(kR,\xi/(\Lambda R)),
\end{align}
where $\kk=k(\cos\theta_k,\sin\theta_k)$ in polar coordinates, and the remaining radial integral is a special case of the integrals
\begin{align}\label{Tndef}
    T_n(a,b)&\equiv \int_0^1 dq\;J_1(aq) q^n\sqrt{\frac{1-q^2}{q^2+b^2}},
\end{align}
that are easily evaluated numerically. 
Note that the angular behavior of \eref{uvka} shows that a vortex in position space is also a vortex in momentum space. We can use \eeref{uvka} to construct the velocity power spectrum for $\psi_{\mathrm{TF}v,a}(\rr)$. Rotational symmetry means that $\varepsilon_\text{kin}^i(k)=2\pi k (m/2)|\tilde{\uu}^i(\kk)|^2$, giving the semi-analytic velocity power spectral density
\begin{align}\label{etfv}
    \varepsilon_{\text{kin,TF}v,a}^i(k)&=\bar\epsilon kR T_1(kR,\xi/(\Lambda R))^2,
\end{align}
where the energy unit is again \eeref{epsbar}.
As a basic test of the spectral analysis, in Appendix \ref{app:nes} we show the spectrum calculated using \eeref{specdens}, compared with the analytical result \eeref{etfv}. The results are identical over all scales. \fref{fig3}(a) we compare \eeref{etfv}, with the GP vortex state. The two agree well at scales less than the system size; in the IR the numerical velocity power spectrum turns over due to the finite system size. The semi-analytical result \eref{etfv} is very close to the numerical result over the entire scale range, apart from a small deviation in the range $k\xi\sim 1$, due to the difference in vortex core shape.
In addition to the different IR behavior, the trapped system also shows oscillations on a scale $k\sim 2\pi/R$, that are not evident in the homogeneous velocity power spectra. Such oscillations also occurred in the spectral density shown in \fref{fig1} for the pure TF wavefunction and are due to the sharp circular boundary at $r=R$ in combination with cylindrical  symmetry. 

Velocity power spectra can also be easily mapped to velocity two-point correlations. Two-point correlations of vorticity provide a useful characterization of point vortices~\cite{novikov_dynamics_1975,skaugen_origin_2017}. Here we calculate a closely related measure accessible directly from the wavefunction of a compressible quantum fluid: the autocorrelation of the incompressible, density weighted velocity field $\uu^i$ defined via Helmholtz decomposition in Section~\ref{sec:1C}. From the general expression \eeref{Guvangle2d}, the particular definition here is
\begin{align}
g_\mathrm{kin}^i(r)&\equiv \frac{1}{C[\uu^i,\uu^i](\mathbf{0})}\int_0^{2\pi}\frac{d\theta}{2\pi}\; C[\uu^i,\uu^i](\rr)
\end{align}
where as usual $C[\uu,\vv](\rr)$ is the two-point correlation \eeref{Cdef}. To use the angle-averaged Wiener-Khinchin theorem, \eeref{wksum}, we use $\langle\uu||\vv\rangle(k)=\varepsilon^i_\textrm{kin}(k)$ in \eeref{Guv}, and with \eeref{Laminvdef} and \eeref{Guvangle2d}, this is simply

\begin{align}\label{gkiexp}
    g_\mathrm{kin}^i(r)&=\frac{1}{E^i_\textrm{kin}}\int_0^\infty dk\; J_0(kr)\varepsilon^i_\textrm{kin}(k).
\end{align}
This integral can be evaluated very accurately due to: (i) the flexibility of our $k$ grid in constructing $\varepsilon^i_\textrm{kin}(k)$; (ii) the behavior in the scale range  $k\xi\gg 1$ is determined by the vortex core structure $\varepsilon^i_\textrm{kin}(k)\sim k^{-3}$, and hence the integral of incompressible energy always converges. 

The angle-averaged two-point correlator of the incompressible velocity is shown in \fref{fig3}(b). The velocity field decorrelates over a scale of order $R$, becoming weakly anticorrelated at long range due the circulation of the fluid. 
\subsection{Central vortex: total kinetic energy}
We now consider a more physical state consisting of a GPE ground state with a single central vortex imprinted in density and phase. The state $\psi_v$ includes quantum pressure from the condensate edge, and the compressible vortex core created using the exact GP numerical solution for the same background density as occurs in the trap centre. To make analytical progress, we also approximate it using $\psi_{va}(\mathbf{r})$, defined in Eq.~(\ref{tfav}), which treats the state as a TF background density with a rational function ansatz for the vortex core.  

For the full kinetic spectral density of $\psi_{\text{TF}v}(\rr)$, we Fourier transform the wavefunction (see Appendix \ref{app:tf}) to find 
the approximate spectral energy density 
\begin{align}\label{finitecore}
    e_{\text{kin,TF}v,a}(k)&= \bar \epsilon(kR)^3T_2(kR,\xi/(\Lambda R))^2.
\end{align}

In \fref{fig3}(c) we compare the semi-analytical energy spectral density with the results for the TF$v$ and GP wavefunctions. We observe good agreement in the infrared, and the numerical TF$v$ and analytical results agree well over a wide range of $k$. Departure from the GPE spectral density is evident for $k \gg 1/R$. The most obvious contrast with \fref{fig1}(a) is that every second peak is suppressed due to vortex lifting the angular degeneracy. Since the vortex core used to construct the TF wavefunction numerically is the true GPE core, the UV departure is due to the extra kinetic energy at the outer edge of the condensate, as also observed in \fref{fig1}(a). 

\subsection{Central vortex: kinetic energy decomposition}
We now decompose the kinetic energy of the central vortex state into compressible, incompressible, and quantum pressure components. The central vortex state has formally zero compressible kinetic energy. Applying the Helmholtz decomposition to a single vortex confirms that the density weighted velocity field $\uu(\rr)\equiv \uu^i(\rr)$ is entirely incompressible~\cite{bradley_energy_2012}. Moreover, for a central vortex in a cylindrical trap density and phase gradients are everywhere orthogonal, so $\uu^i(\rr)\cdot\uu^q(\rr)=0$. This does not preclude a finite coupling between quantum pressure and incompressible energy, $e_\text{kin}^{iq}(k)$, that can acquire negative values and so redistributes energy between different scales as seen in \fref{fig3}(d). Nevertheless, as shown in \eeref{Eabint}, the \emph{total} contribution to the energy found by integrating $e_\text{kin}^{iq}(k)$ is always zero, and the total energy density computed directly from the wavefunction without decomposition, $e_{\mathrm{kin},v}(k)$, agrees exactly with the sum of incompressible and quantum pressure energy densities, $e_{\mathrm{sum},v}(k)$. 

The quantum pressure terms depart for $kR\gg 1$, as should be expected from the sharp feature at the TF boundary, $r=R$. The construction of the states means that the vortex core plays no role in this difference, as in both states the core is the exact GPE core.  

The essential role of the coupling energy between incompressible and quantum pressure components may be observed in \fref{fig3}(d). The coupling term $e_\text{kin}^{iq}(k)$ serves to redistribute energy, and must be included in the sum, \eeref{ekindens}, in order to correctly reproduce $e_\text{kin}(k)$. In \fref{fig3}(c) and \fref{fig3}(d) rotational symmetry breaking appears to suppress every second peak, relative to \fref{fig1}(a). In \fref{fig3}(d) we see the coupling term $e_\text{kin}^{iq}(k)$ redistributing energy between peaks creating the suppression and enhancement of alternate peaks. 

\subsection{Vortex Distributions}
\begin{figure}[!t]
    \centering
    \includegraphics[width=\columnwidth]{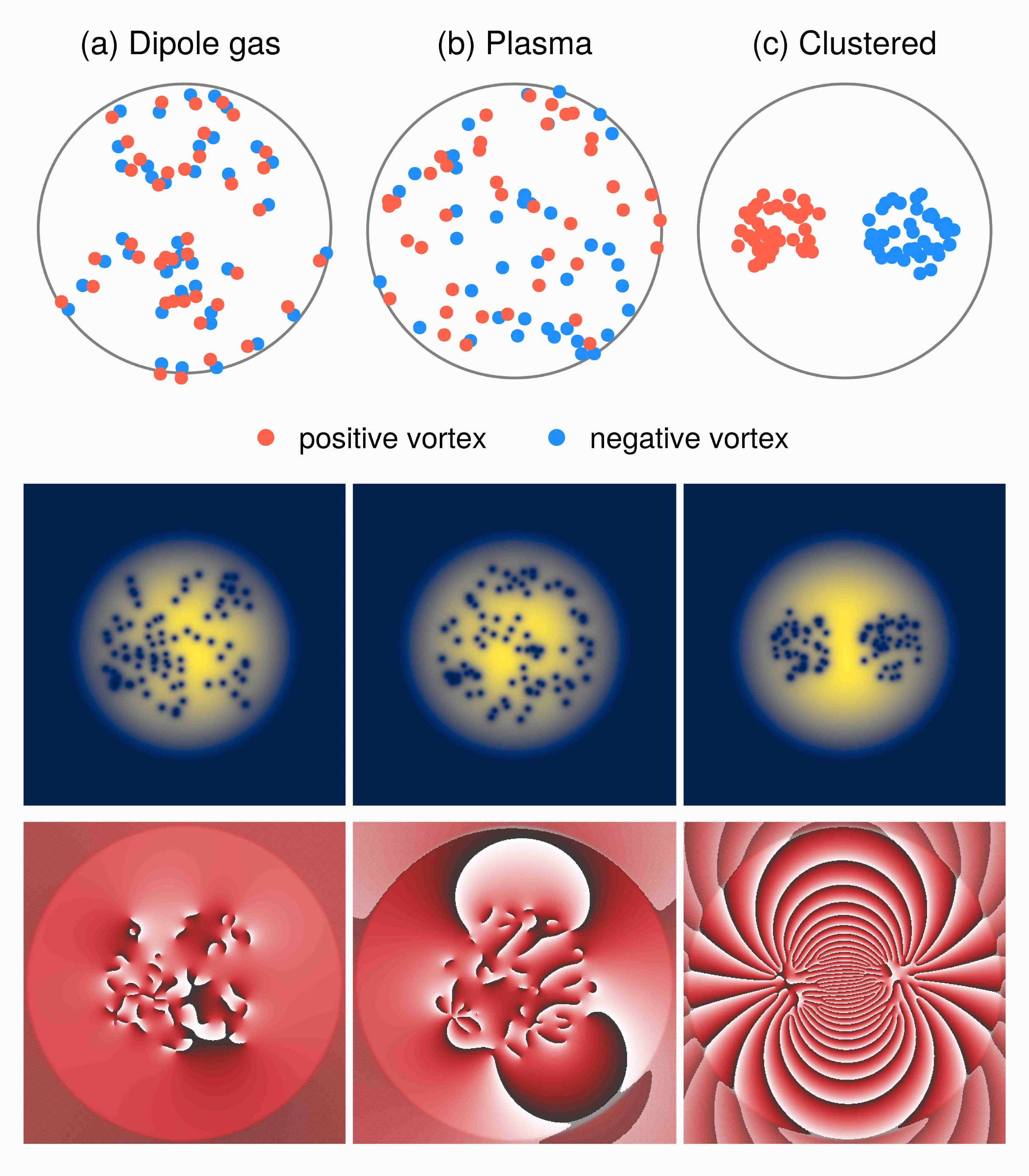}
    \caption{Top: Schematic of regimes for neutral planar quantum vortex distributions in a disc domain $r<R_\circ$. (a) \emph{Dipole gas} with dipole centroids uniformly distributed on the disc, angles uniform on $[0,2\pi)$, and fixed dipole length $d\ll R_\circ$; (b) \emph{Plasma} with vortices uniformly distributed on the disc of radius $R_\circ$; (c) \emph{Clustered} distribution where vortices of each sign are distributed in smaller discs of radius $R_c<R_\circ$, with centers separated by $D$. For each regime we show the atomic density (middle) and quantum phase (bottom) of the wavefunction for a single sample. See text for parameters; spatial regions and colormaps are defined in Fig. \ref{fig2}.}
    \label{fig4}
\end{figure}

We calculate the energy spectral densities and velocity power spectra for neutral distributions in the same harmonically trapped system considered in the previous section. We sample three distributions shown schematically in \fref{fig4}(a-c), with representative samples after imprinting on the GPE ground state: 
\begin{enumerate}[(a)]
    \item Vortex dipoles distributed uniformly, with random orientations, within a disc $r< R_\circ =0.7R$. The dipole centroids are uniformly distributed on the disc, and the dipole length $d=5.48\xi$ corresponds to a low energy \emph{dipole gas}, \fref{fig4}(a).  
    \item Uniformly distributed vortices, within the same disc $r< R_\circ$, and  uncorrelated with respect to both sign and position. In a uniform BEC this uncorrelated state, referred to as the vortex \emph{plasma} has a universal $k^{-1}$ velocity power spectrum~\cite{kusumura_energy_2013}, as seen in \fref{fig4}(b). 
    \item Sign-polarized and sampled uniformly within two smaller discs (radius $R_c=0.3R$) separated by distance $D=0.8R=34\xi$, corresponding to a high-energy \emph{clustered} state, \fref{fig4}(c).
\end{enumerate}

Each of these states is sampled using suitably scaled and shifted uniform random variates.
We imprint vortices using the numerically exact (homogeneous) GPE core solution, using local GPE healing length, and the ideal quantum phase for a vortex in a homogeneous background density. Individual sample wavefunctions in the three regimes are shown in \fref{fig4}. 

\begin{figure}[!t]
    \centering
    \includegraphics[width=\columnwidth]{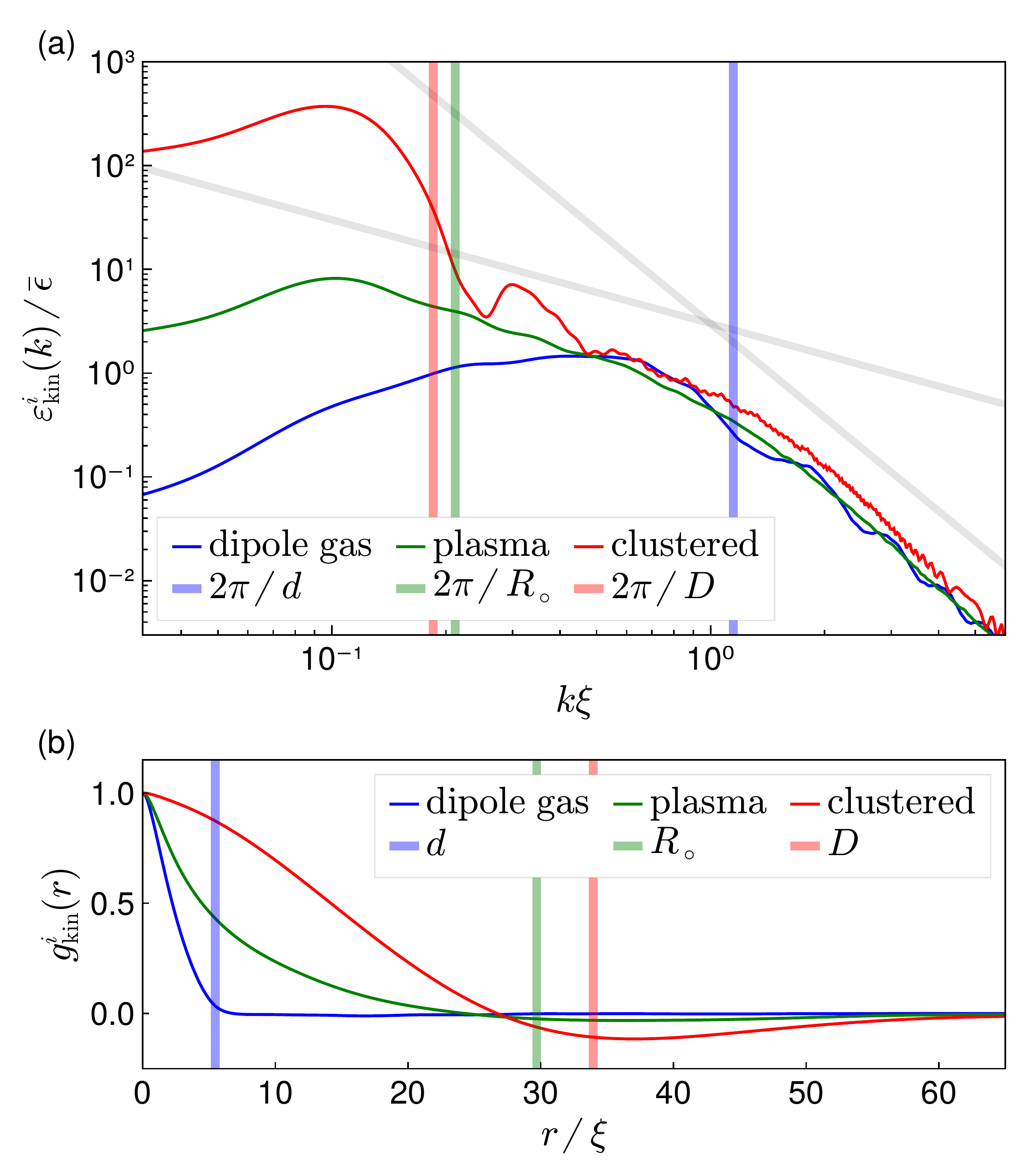}
    \caption{(a) Velocity power spectra of the incompressible velocity field for dipole gas, plasma, and clustered vortex distributions. The gray lines show $k^{-1}$ and $k^{-3}$ power laws for comparison; (b) Angle-averaged autocorrelation of the incompressible velocity, \eeref{gkiexp}, for dipole gas, plasma, and clustered distributions.}
    \label{fig5}
\end{figure}

We calculate velocity power spectrum, \eeref{ecspecdens}, once again using the convolution expression \eref{Ekkic}. 
The velocity power spectra for the incompressible velocity, averaged over $N_s=50$ samples, are presented in \fref{fig5}(a). The notable features are (i) the $k^{-3}$ ultraviolet region, associated with the core of a single planar quantum vortex core~\cite{krstulovic_comment_2010-1,nowak_superfluid_2011,bradley_energy_2012}; (ii) The extended $\sim k^{-1}$ region for the plasma phase \cite{nowak_superfluid_2011,nowak_nonthermal_2012,kusumura_energy_2013}  (iii) Significant increasing in the velocity power spectrum in the infrared as the state proceeds from dipole gas to clustered. 

Starting from the numerically accurate spectra shown in \fref{fig5}(a), we compute the velocity autocorrelation using the exact mapping \eref{wksum}, to give the results shown in \fref{fig5}(b). 
This provides a compact representation of incompressible velocity correlations for the system. 
The dominant velocity correlation length scales can be seen clearly: the dipole gas spatial correlations decay over scale $\xi\ll d\ll R_\circ $; the plasma phase has velocity correlations decaying over the scale $\sim R_\circ$, the size of the distribution; the clustered phase develops velocity anticorrelations peaked at the cluster separation scale $\sim D$.

\section{Conclusions}\label{sec:conc}

We have developed a spectral analysis for compressible quantum fluids that includes all quantum phase information and clearly distinguishes between velocity power spectral densities and energy spectral densities. Focusing on the example of a Gross-Pitaevskii fluid, we have obtained well-resolved and numerically accurate angle-averaged quantities in momentum space, and tested against analytical results for a central vortex in a 2D harmonic trap. Our treatment provides a decomposition into compressible, incompressible, and quantum pressure contributions for energy spectral densities and velocity power spectra. The latter are standard measures of quantum turbulence~\cite{bradley_energy_2012,nore_kolmogorov_1997} that have a classical origin and explain energy fluxes in incompressible classical fluids. In compressible quantum fluids the former are required for a full account of energy transport, and e.g. rigorous definition of energy fluxes. A complete treatment of energy transport in scale space for compressible quantum fluids in a range of flow scenarios offers an interesting future direction. 

The semi-analytically tractable two-dimensional vortex state provides a strong test of spectral analysis, and allowed verification of local additivity of energy spectral densities. High resolution velocity power spectra also allow convenient evaluation of angle-averaged velocity autocorrelation functions, providing sufficient information in the infrared to give an accurate reconstruction in position space. Representative vortex distributions show the qualitative value of such correlations as a system-wide representation of significant correlation lengths for the dipole gas, plasma, and clustered phases of neutral quantum vortices. It would be very interesting to apply these measures to a broader class of vortex or phonon dominated flows. 

We have focused on kinetic energy spectra due to their importance in quantum turbulence studies. However, the total kinetic energy spectrum is essentially equivalent (up to a polynomial factor) to the occupation number per mode $n(k)$, the central quantity in the study of non-thermal fixed points~\cite{nowak_universal_2014,mathey_anomalous_2015}. High resolution spectral analysis may also address a similar challenge faced in computing the superfluid fraction at finite temperature via momentum correlations~\cite{foster_vortex_2010}. 
Our approach also offers a pathway for robust calculation of energy fluxes in quantum fluids, central to measures of turbulent cascades~\cite{billam_spectral_2015,fujimoto_direct_2016,numasato_direct_2010,griffin_energy_2022}. The spectral analysis developed here offers a quantitative path to address the question: what is truly quantum in quantum turbulence?

\begin{acknowledgments}
    AB thanks Brian P. Anderson and Matthew T. Reeves for stimulating discussions. AB acknowledges support from the Marsden Fund with grant No. UOO1726, and the Dodd-Walls Centre for Photonic and Quantum Technologies. XY acknowledges the support from NSAF with grant No. U1930403 and NSFC with grant No. 12175215. 
\end{acknowledgments}

\appendix
\section{Helmholtz decomposition}\label{app:helm}
The decomposition may be carried out in momentum space by the projection parallel and orthogonal to the radial unit vector $\hat\kk\equiv\kk/|\kk|$. Defining 
\begin{align}
    \tilde{\uu}(\kk)&\equiv \frac{1}{(2\pi)^{d/2}}\int d^d\rr\;e^{-i\kk\cdot\rr}\uu(\rr)=\FT{\mathbf{u}},
\end{align}
for a general vector field, the decomposition is written in momentum space as
\begin{align}
    \tilde{u}^c_i(\kk)&=\sum_j\frac{k_ik_j}{k^2}\tilde{u}_j(\kk),\notag\\
    \tilde{u}^i_i(\kk)&=\sum_j \left(\delta_{ij}-\frac{k_ik_j}{k^2}\right)\tilde{u}_j(\kk),
\end{align} 
where $\delta_{ij}$ is the Kronecker-delta, $\uu^{(i,c)}(\rr)=\IFT{\tilde\uu^{(i,c)}}$, and subscripts refer to cartesian components $i\in \{1,\dots,d\}$. The incompressible or longitudinal component is associated with quantum vortices. The compressible or transverse component is associated with density variations, including Bogoliubov quasiparticles in the small fluctuation limit.

\section{Angle-averaged Wiener-Khinchin theorem}
\label{app:wkt}

The angle-averaged Wiener-Khinchin theorem is most clearly and generally formulated by considering the inner product of two vector fields $\uu(\mathbf{r})$ and $\vv(\mathbf{r})$ in $n$ dimensions, expressed in Fourier space as
\begin{align}\label{ip1}
    \langle \uu|\vv\rangle&=\int d^d\kk\;\langle \uu|\kk\rangle\langle\kk|\vv\rangle.
\end{align}
We emphasize that the bra-ket notation used here is convenient for expressing basis changes, but should only be interpreted in the pure-state sense with respect to the wavefunction, as is standard in mean-field theory; our approach can be generalized beyond mean-field theory via truncated Wigner or other semi-classical quantum Monte-Carlo methods~\cite{blakie_dynamics_2008}. 
We can write \eeref{ip1} more usefully by expanding the Dirac-delta as 
\begin{align}
    \ip{\uu}{\vv}&=\int d^d\kk\;\int d^d\rr \int d^d\rr'\langle \uu|\rr'\rangle\langle\rr'|\kk\rangle\langle \kk|\rr\rangle\langle \rr|\vv\rangle.
\end{align}
The Dirac-delta function can be evaluated in Fourier space by integrating over the relevant $k$-space (solid) angle to give
\begin{align} \label{deltagen}
    \delta^{(d)}(\rr-\rr')&=\int d^d\kk\; \langle\rr'|\kk\rangle\langle \kk|\rr\rangle\notag\\
    &=\int_0^\infty dk\; \Lambda_d(k,|\rr-\rr'|),
\end{align}
where 
\begin{align}\label{Lamdef}
    \Lambda_d(k,r)&\equiv
    \begin{cases}
        \frac{1}{2\pi}\;kJ_0(kr),&\text{for } d=2,\\
        \frac{1}{2\pi^2}\;k^2\sinc{(kr)},&\text{for } d=3,
    \end{cases}
\end{align}
and $\sinc{(x)}\equiv \sin{(x)}/x$.
We can thus write the inner product as
\begin{align}
    \langle \uu|\vv\rangle&=\int_0^\infty dk\;\langle\uu||\vv\rangle(k),
\end{align}
where the \emph{spectral density} of the inner product is defined as
\begin{align}\label{specdens0}
    \langle\uu||\vv\rangle(k)&\equiv\int d^d\rr\int d^d\rr'\;\Lambda_d(k,|\rr-\rr'|)\langle \uu|\rr'\rangle\langle \rr|\vv\rangle.
\end{align}
The change of variables 
\begin{align}\label{vchange}
\mathbf{R}&=\tfrac{1}{2}(\rr+\rr'),\quad\x=\rr-\rr',
\end{align}
with unit Jacobian, gives 
\begin{align}\label{specdens}
    \langle\uu||\vv\rangle(k)&\equiv\int d^d\x\;\Lambda_d(k,\x)C[\uu,\vv](\x),
\end{align}
where 
\begin{align}\label{Cdef}
    C[\uu,\vv](\x)&\equiv\int d^d\mathbf{R}\;\ip {\uu}{\mathbf{R}-\x/2}\ip{\mathbf{R}+\x/2}{\vv},
\end{align}
is the two-point correlation function of the vector fields, \eeref{2pc}. Equation \eref{specdens} is the main result of this appendix, and forms the basis of an accurate spectral analysis developed in this work. We now use this formulation to develop a formal mapping between spectral density and system averaged two-point correlations. 

Two special cases are worth emphasizing. First, at $\mathbf{x}=\mathbf{0}$, the correlation reduces to the vector inner product $C[\uu,\vv](\mathbf{0})\equiv\ip{\uu}{\vv}$. Second,  for any two vector fields that are orthogonal everywhere in momentum space the correlation vanishes, $C[\uu,\vv](\mathbf{0})\equiv 0$, at $\mathbf{x}=\mathbf{0}$. We prove this property in \aref{app:coupling}, and use it in \sref{decompE} where we analyse the spectral energy densities. 

In each spatial dimensionality there is a well defined inversion operation that transforms the spectrum to an angle-averaged two-point correlation function in position space. The inverse is defined as
\begin{align}\label{Guv}
    G_{uv}(r)\equiv\int_0^\infty dk\;\Lambda_d^{-1}(k,r)\langle\uu||\vv\rangle(k),
\end{align}
with inverse kernel
\begin{align}\label{Laminvdef}
    \Lambda_d^{-1}(k,r)&\equiv\begin{cases}
        J_0(kr)&\text{for } d=2,\\
        \sinc{(kr)}&\text{for } d=3,
    \end{cases}
\end{align}
as follows immediately from the closure relations for Bessel and sinc functions
\begin{align}
    \int_0^\infty dk\;kJ_0(kr)J_0(kr')&=\frac{1}{r}\delta(r-r'),\\
    \int_0^\infty dk\;k^2\sinc(kr)\sinc(kr')&=\frac{\pi}{2r^2}\delta(r-r')\nonumber\\
    &+\frac{\pi}{2r^2}\delta(r+r').
\end{align}
Using the closure relations and the spectral density \eref{specdens}, in cylindrical and spherical coordinates, we evaluate \eref{Guv} to arrive at the angle-averaged correlation functions in the form
\begin{align}\label{Guvgen}
    G_{uv}(r)&=
        \frac{1}{\Omega_d}\int d^d\rr' \delta(r-|\rr'|)\; C[\uu,\vv](\rr'),
\end{align}
where $\Omega_d=\int d\Omega_d$ is the total solid angle in dimension $d$. In detail, for $d=2,3$ we have 
\begin{align}\label{Guvangle2d}
    G_{uv}(r)&=
        \frac{1}{2\pi}\int_0^{2\pi}d\theta\; C[\uu,\vv](r\cos\theta,r\sin\theta),
\end{align}
and
\begin{align}\label{Guvangle3d}
    G_{uv}(r)&=\frac{1}{4\pi}\int_0^{2\pi}d\phi\int_0^\pi d\theta\;\sin\theta\notag\\
    &\times C[\uu,\vv](r\sin\theta\cos\phi,r\sin\theta\sin\phi,r\cos\theta),
\end{align}
respectively. This establishes the formal connection between the inverse Eq.~\eref{Guv} and the angle-averaged two-point correlation function in position space. The two-point correlation may be computed conveniently by first calculating the power spectral density \eref{specdens} with high resolution, and then using the definition \eref{Guv} to transform to position space. The value of $G_{uv}$ at $r=0$ stems from the inner product, and does not provide any useful information about correlations for $r\neq 0$. We can choose the normalization $g_{uv}(0)=1$, in which case we work with 
\begin{align}\label{gnormed}
    g_{uv}(r)&\equiv \frac{G_{uv}(r)}{C[\uu,\vv](\mathbf{0})}=\frac{G_{uv}(r)}{G_{uv}(0)}. 
\end{align}
In summary, for any two vector fields $\uu$,$\vv$, we use the definitions Eqs.~\eref{specdens}, \eref{Lamdef}, and \eref{Guv}, \eref{Laminvdef}, \eref{Guvgen} to  construct an angle-averaged power spectrum in $k$-space, and a corresponding angle-averaged two-point correlation function in position space 
\begin{align}\label{wksum}
    (\uu,\vv)\longrightarrow \langle \uu||\vv\rangle(k)\longleftrightarrow g_{uv}(r),
\end{align}
where the first mapping, \eeref{specdens}, is a Fourier transform followed angular integration in $k$ space, and the second, \eeref{Guv}, is the Fourier relation between the $k$ and $r$ variables. Our analysis amounts to an explicitly angle-averaged formulation of the standard Wiener-Khinchin theorem linking power spectra with correlations. 

\section{Vanishing of coupling term contributions to the total energy}\label{app:coupling}
We show that the coupling terms integrate to zero. 
For $e_\text{kin}^{ic}(k)$, using \eeref{deltagen} gives
\begin{align}\label{einczero}
    \int_0^\infty dk\;e_\text{kin}^{ic}(k)&=\frac{m}{2}\int d^d\x\;\delta^{(d)}(\x)\Big\{C[\ww^i,\ww^c](\x)\notag\\
    &+C[\ww^c,\ww^i](\x)\Big\}\notag\\
    &=m\textrm{Re}\{ C[\uu^i,\uu^c](\mathbf{0})\}.
\end{align}
Although the fields are not in general orthogonal in position space, in momentum space they are always orthogonal due to the Helmholtz decomposition. For the special case of an incompressible and compressible field, the correlation in \eref{einczero} is identically zero: from \eref{Cdef}, using the standard completeness and orthogonality relations for position and momentum eigenstates, we have
\begin{align}
    C[\uu^i,\uu^c](\mathbf{0})&=\int d^d\mathbf{R}\;\ip {\uu^i}{\mathbf{R}}\ip{\mathbf{R}}{\uu^c}\notag\\
    &=\int d^d\mathbf{R}\;\int d^d\mathbf{k}\;\int d^d\mathbf{k}'\notag\\
    &\times\ip{\uu^i}{\mathbf{k}}\ip{\mathbf{k}}{\mathbf{R}}\ip{\mathbf{R}}{\mathbf{k}'}\ip{\mathbf{k}'}{\uu^c}\notag\\
    &=\int d^d\mathbf{k}\;\int d^d\mathbf{k}'\ip{\uu^i}{\mathbf{k}}\ip{\mathbf{k}}{\mathbf{k}'}\ip{\mathbf{k}'}{\uu^c}\notag\\
    &=\int d^d\mathbf{k}\;\ip{\uu^i}{\mathbf{k}}\ip{\mathbf{k}}{\uu^c}\notag\\
    &=\int d^d\mathbf{k}\;\uu^i(\mathbf{k})^*\cdot\uu^c(\mathbf{k})=0,
\end{align}
since these fields are orthogonal in momentum space by construction.
For $e_\text{kin}^{iq}(k)$ we have
\begin{align}
    \int_0^\infty dk\;e_\text{kin}^{iq}(k)&=\frac{m}{2}\int d^d\x\;\delta^{(d)}(\x)\Big\{C[\ww^i,\ww^q](\x)\notag\\
    &+C[\ww^q,\ww^i](\x)\Big\}\notag\\
    &=m\textrm{Im}\{C[\uu^i,\uu^q](\mathbf{0})\}= 0.
\end{align}
Here the result follows from the fact that the correlator of real-valued fields is always real valued (the fields are not strictly orthogonal in any space). $\int_0^\infty dk\;e_\text{kin}^{cq}(k)=0$ follows similarly. 

\section{Numerical convergence of spectra}\label{app:nes}
\begin{figure}[!t]
    \centering
    \includegraphics[width=\columnwidth]{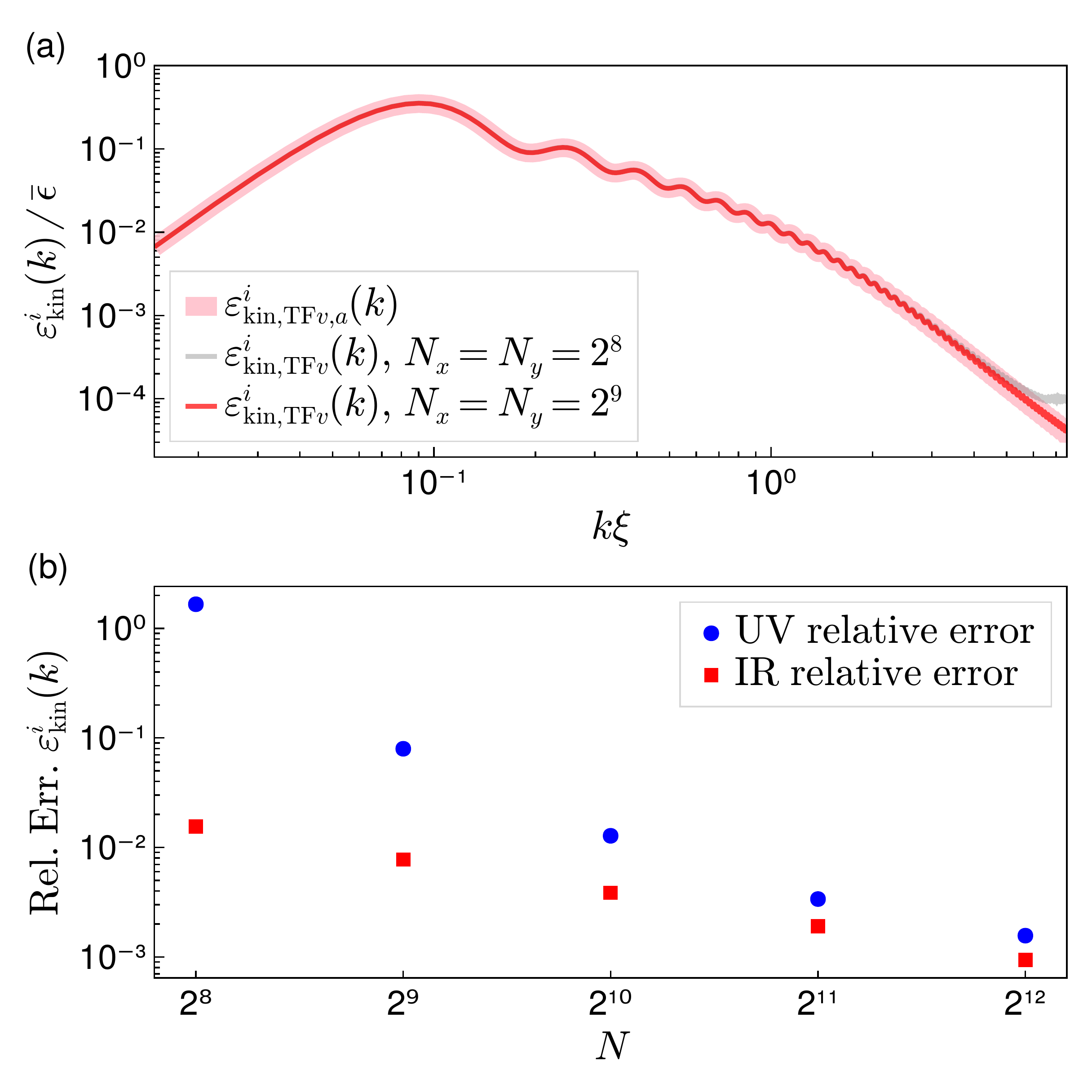}
    \caption{Thomas-Fermi state with ansatz core: (a) Velocity power spectra showing the analytical result with numerical spectra for two different grid point densities on the same spatial domain shown in \fref{fig2}; (b) relative error in the IR and UV regions as a function of grid point density on the same domain, over the interval of $k$ shown in (a). The reported UV error is also the maximum relative error for all $k$; The IR error is the constant relative error that is rapidly approached for $k\xi\lesssim 1$.\label{fig6}}
\end{figure}

One test of numerical accuracy would be to compare our results with approximate spectra computed using binning into annular regions of $k$-space. However, since we can compute the spectrum at any $k$ values using \eeref{specdens}, we instead examine the maximum relative error over a range of $k$, as tested against the analytical result \eref{etfv}. This offers a much more exacting test of the numerical spectral analysis. We show that the numerically computed spectrum for $\psi_{\text{TF}v}(\rr)$ converges to the analytical spectrum $\varepsilon_{\text{kin,TF}v,a}^i(k)$ for all $k$ including in UV region of the vortex core. In \fref{fig6}(a) we show the analytical velocity power spectrum, compared with the numerically computed spectrum for $N_x=N_y=N=256$ points and $N=512$ points. Over most of the scale range the error is small, and it only becomes significant inside the vortex core where $k\xi\gg 1$, as seen for $N=256$. In \fref{fig6}(b) we show the relative error in the IR regime, and the UV regime (also the maximum over the entire range of $k$). The relative error is always small in the IR. It convergences uniformly with increasing $N$ in the UV region, despite the amplification in the vortex core due to the very small values of $\varepsilon_{\text{kin,TF}v}^i(k)$. For $N\geq 512$, even the UV relative error is small, corresponding to choosing a grid with at least 4 points per healing length. 2 points per $\xi$ achieves very good spectral resolution for $k\xi\lesssim 1$.

\section{Thomas-Fermi integrals}\label{app:tf}
From the Thomas-Fermi wavefunction for the ground state,~\eeref{tfwave}, we compute Fourier transforms of the GPE energy terms using cylindrical symmetry
\begin{align}
    E_\text{pot}&=\int d^2\rr\; V(\rr)|\psi|^2=\int d^2\kk\;|f(\kk)|^2,\\
    E_\text{int}&=\frac{g}{2}\int d^2\kk\;|\tilde n(\kk)|^2,
\end{align}
where 
\begin{align}
    \phi(\kk)&=\frac{1}{2\pi}\int d^2\rr\;e^{-i\kk\cdot\rr}\psi(\rr) 
    =\sqrt{\frac{\mu}{g}}R^2I_1(kR),\\
    f(\kk)&\equiv\int \frac{d^2\rr}{2\pi}\; e^{-i\kk\cdot\rr}\sqrt{V(\rr)n(\rr)}=\sqrt{\frac{m\omega^2\mu}{2g}}R^3I_2(kR),\\
    \tilde n(\kk)&\equiv\frac{1}{2\pi}\int d^2\rr\;e^{-i\kk\cdot\rr}n(\rr)=n_0R^2 I_3(kR),
\end{align}
and the integrals are 
\begin{align}
    I_1(a)&=\int_0^1 dq\;J_0(aq)q\sqrt{1-q^2}=\frac{\sin{a}-a\cos{a}}{a^3},\\
    I_2(a)&=\int_0^1 dq\; q^2\sqrt{1-q^2}J_0(aq)\notag\\
    &=\frac{\pi}{2a^2}\left[aJ_0(a/2)-2J_1(a/2)\right]J_1(a/2),\\
    I_3(a)&=\int_0^1 dq\;J_0(aq)q(1-q^2)=\frac{2J_2(a)}{a^2}.
\end{align}
Using these integrals, we arrive at the expressions \eref{Etfkin}-\eref{Etfpot}.

To compute the velocity power spectrum for $\psi_{\text{TF},v}(\rr)$, we want to find the Fourier transform of the density-weighted velocity field $\tilde{u}^i(\kk)$, defined in \eref{uvr}. In cylindrical polar coordinates $\kk=k(\cos\theta_k,\sin\theta_k)$, $\kk\cdot \rr=rk\cos(\theta-\theta_k)$. and we have  
\begin{align}
    \tilde{\uu}^i(\kk)&=\pm\frac{\sqrt{n_0}\hbar}{m}\int_0^R dr\;\frac{r\sqrt{1-r^2/R^2}}{\sqrt{r^2+(\xi /\Lambda)^2}}\notag\\
    &\times\frac{1}{2\pi}\int_0^{2\pi} d\theta\; e^{-ikr\cos(\theta-\theta_k)}(-\sin\theta,\cos\theta).
\end{align}
Using the Bessel integrals
\begin{align}\label{besselint}
\frac{1}{2\pi}\int_0^{2\pi}d\theta\;e^{-i\beta\cos(\theta-\phi)}\begin{cases}
    \cos\theta \\
    \sin\theta  
    \end{cases}&=-iJ_1(\beta)\begin{cases}
        \cos\phi \\
        \sin\phi 
        \end{cases},
\end{align}
and changing variables to $q=r/R$, we arrive at \eref{uvka}. 

We complete this appendix with a derivation of the kinetic energy density of a TF state containing a vortex, \eeref{finitecore}. Using the definition \eeref{ekin_all}, we start by Fourier transforming the TF$v$ wavefunction 
\begin{align}\label{psivk}
    \phi_{va}(\kk)&=\frac{1}{2\pi}\int d^2\rr\;e^{-i\kk\cdot\rr}\psi_{va}(\rr)\notag\\
    &=\sqrt{n_0}\int_0^R dr\;\frac{r^2}{\sqrt{r^2+(\xi /\Lambda)^2}}\sqrt{1-\frac{r^2}{R^2}}\notag\\
    &\times\frac{1}{2\pi}\int_0^{2\pi} d\theta\; e^{-ikr\cos(\theta-\theta_k)+i\theta}.
\end{align}
Using \eref{besselint} and changing variables to $q=r/R$ gives
\begin{align}
    \phi_{va}(\kk)&=-i\sqrt{n_0}R^2\int_0^1 dq\frac{q^2\sqrt{1- q^2 }}{\sqrt{q^2+(\xi /\Lambda R)^2}}J_1(kRq)\notag\\
    &=-iR^2\sqrt{n_0}T_2(kR,\xi/(\Lambda R)),
\end{align}

Using cylindrical symmetry, we have 
\begin{align}
    \varepsilon_{\mathrm{kin,TV}v,a}(k)&=2\pi k (\hbar^2k^2/2m)|\phi_{\text{TF}v,a}(\kk)|^2,
\end{align}
and we arrive at \eeref{finitecore}.


\providecommand{\noopsort}[1]{}

\end{document}